\documentclass[useAMS,usenatbib]{mn2e}

\newcommand{\beq}{\begin{equation}}
\newcommand{\eeq}{\end{equation}}
\newcommand{\bdm}{\begin{displaymath}}
\newcommand{\edm}{\end{displaymath}}
\newcommand{\bea}{\begin{eqnarray}}
\newcommand{\eea}{\end{eqnarray}}
\newcommand{\bt}{\begin{tabular}}
\newcommand{\et}{\end{tabular}}

\usepackage{graphicx}
\usepackage{times}%
\usepackage{psfig,epsfig} 
\usepackage{natbib}

\newcommand{\lesssim}{\la} 
\newcommand{\fnl}{f_{\rm{NL}}}

\title[The density field in non-Gaussian models]
{The mass density field in simulated non-Gaussian scenarios}
\author[M. Grossi et al.] 
{M. Grossi$^{1}$, 
E. Branchini$^{2}$,
K. Dolag$^{1}$,
S. Matarrese$^{3,4}$, 
L. Moscardini$^{5,6}$\\
$^1$ Max-Planck Institut fuer Astrophysik,
Karl-Schwarzschild Strasse 1, D-85748 Garching, Germany
(margot,kdolag@mpa-garching.mpg.de)\\
$^{2}$ Dipartimento di Fisica, Universit\`a di Roma TRE,
via della Vasca Navale 84, I-00146, Roma, Italy
(branchin@fis.uniroma3.it)\\
$^{3}$ Dipartimento di Fisica, Universit\`a di Padova,
via Marzolo 8, I-35131, Padova, Italy
(sabino.matarrese@pd.infn.it)\\
$^{4}$ INFN, Sezione di Padova,
via Marzolo 8, I-35131, Padova, Italy\\
$^{5}$ Dipartimento di Astronomia, Universit\`a di Bologna,
via Ranzani 1, I-40127 Bologna, Italy 
(lauro.moscardini@unibo.it)\\
$^{6}$ INFN, Sezione di Bologna, viale Berti Pichat 6/2,
I-40127 Bologna, Italy\\
}

\date{}
\begin{document}

\date{Accepted ???. Received ???; in original form April 2008}

\pagerange{\pageref{firstpage}--\pageref{lastpage}} \pubyear{2008}

\maketitle

\label{firstpage}

\begin{abstract}
  In this work we study the properties of the mass density field in
  the non-Gaussian world models simulated by \cite{grossi2007}.  
  In particular we focus on the one-point density probability
  distribution function of the mass density field in non-Gausian models 
  with quadratic non-linearities quantified by the usual parameter
  $f_{\rm{NL}}$. We find that the imprints of primordial non-Gaussianity 
  are well preserved in the negative tail of the probability function 
  during the evolution of the density perturbation.
  The effect is already noticeable at redshifts as large as 4 and 
  can be detected out to the present epoch. At $z=0$ we find that 
  the fraction of the volume occupied by regions with underdensity 
  $\delta < -0.9$, typical of voids,  is about 1.3 per cent in the Gaussian case
  and increases to $\sim 2.2$ per cent   if $f_{\rm{NL}}=-1000$
  while decreases to $\sim 0.5$ per cent if $f_{\rm{NL}}=+1000$. 
  This result suggests that 
  void-based statistics may provide a powerful method to detect
  non-Gaussianity even at low redshifts which is complementary to
  the measurements of the   higher-order moments of the probability
  distribution function like the skewness or the kurtosis for which 
  deviations from the Gaussian case are detected at the 25-50 per cent level.

\end{abstract}


\section{Introduction}\label{sect:introduction}

The renewed interest on studying deviations from Gaussianity in the
primordial density fields has been recently prompted by new
observations and theoretical progress.

On the theoretical side, inflation-based non-Gaussian (NG) models have
been thoroughly investigated \citep[see, e.g.,][and references
therein]{bartolo2004} and can be used to make robust predictions for
quantities that can be potentially observed like the mass function of
virialized structures \citep{matarrese2000}, their biasing
\citep{matarrese1986,grinstein1986,dalal2007,
matarrese2008}, the topology of the large scale structure
(LSS) of the universe \citep{matsubara2003}, and higher-order
clustering statistics like the bi-spectrum \citep{hikage2006}.

On the observational side, the cosmic microwave background (CMB) has
provided the most stringent constraints so far on the level of
non-Gaussianity. The most recent analysis of the WMAP 5-year
temperature fluctuation maps \citep{komatsu2008} shows that the
primordial NG signal on the very large scales probed by CMB is tiny
(see the discussion in Sect.\ref{sect:nbody}). Yet the latest claims
of a positive detection of NG features in the WMAP 3-year data
reported by \cite{yadav2007} has renewed the interest in NG models,
shifting the focus to smaller scales where NG features can only be
spotted through the analysis of the Large Scale Structure (LSS) of the
universe at much lower redshifts. In this respect, current
observations and future planned experiments will certainly deliver new
information capable of setting new and stringent constraints on the
amount of non-Gaussianity.

In this framework numerical experiments represent the only tool
capable of bridging the gap between theoretical analytic predictions
that usually involve simplifying hypotheses, and LSS-based
observations, where non-linear effects play a non-negligible role.  In
this respect, different groups have performed new, independent
numerical experiments \citep[see,
e.g.,][]{mathis2004,kang2007,grossi2007,dalal2007}.

In this work we use the highest-resolution experiments performed so
far of physically motivated NG models that have been described in
  detail by \citep[][hereafter G07]{grossi2007}.  In that work we have
  studied the evolution of massive dark matter haloes in non-Gaussian
  scenarios quantified by the mass function, and compared it with
  analytic predictions. In a subsequent paper we have used the same
  numerical experiments to investigate the topological properties of
  the mass density field quantified by the Minkowski Functionals
  \citep{hikage2008}.

  Here we focus on the probability distribution function of the mass
  density field [PDF hereafter] which describes the probability that a
  randomly placed cell of specified volume and shape contains a
  specified density.  In absence of primordial non-Gaussianity,
  the PDF has a Gaussian form initially, but develops a positive
  skewness and kurtosis at later times.  Deviations from the original
  Gaussian shape originate from the non-linear evolution of density
  fluctuations and non-linear distortions induced by departures
  from the Hubble flow, if redshifts are used to estimate galaxy
  distances.  In our analysis we will ignore the latter and use  
  N-body simulations to account for non-Gaussianity  induced 
  by non-linear dynamics with the purpose of characterizing 
  the best range of scales and density in which 
  primordial non-Gaussianity dominates over late contributions.
  
  Our approach should be regarded as complementary to 
  the analytic methods that model the evolution of the PDF in a
  Gaussian scenario using either perturbation theory or excursion set
  methods \citep[see, e.g.,][and references
  therein]{lam2008}. Assessing the goodness of these methods is beyond
  the scope of this work. However, as a by-product, we will compare 
  our result to the  lognormal model to evaluate whether this simple analytic 
  prediction is accurate enough to allow measuring non-Gaussianity
  in the PDF or whether more sophisticated models are required. 
  
  Finally, we point out
  that our analysis should be regarded as somewhat ideal since, like
  in majority of the theoretical PDF studies, we focus on the effects
  on the non-linear evolution of density fluctuations ignoring
  not only  redshift-space distortions but also deviations from Gaussianity
  due to  galaxy biasing.
 
This paper is organized as follows. In Section~\ref{sect:nongaussian}
we introduce the NG model considered here. The characteristics of our
N-body simulations are presented in Section~\ref{sect:nbody}, where we
also give details on the method adopted to implement suitable NG
initial conditions.  In Section~\ref{sect:evolution} we illustrate how
primordial non-Gaussianitities affect the redshift evolution of
different quantities like the clustering properties
(Section~\ref{sect:clustering}), the probability distribution function
of density fluctuations (Section~\ref{sect:pdf}) and its high-order
moments (Section~\ref{sect:moments}) We discuss the results and
conclude in Section~\ref{sect:conclusions}.

\section{Non-Gaussian models}\label{sect:nongaussian}

It is possible to model the level of primordial non-Gaussianity
predicted by a large class of models for the generation of the initial
seeds for structure formation (including standard single-field and
multi-field inflation, the curvaton and the inhomogeneous reheating
scenarios) using a quadratic term in the Bardeen's gauge-invariant
potential $\Phi$, which, on scales much smaller than the Hubble
radius, reduces to minus the usual peculiar gravitational potential:
\begin{equation}
\label{FNL}
\Phi = \Phi_{\rm L} + f_{\rm{NL}} \left(\Phi_{\rm L}^2 - 
\langle\Phi_{\rm L}^2\rangle \right) \; .
\end{equation}
In the previous equation $\Phi_{\rm L}$ represents a Gaussian random
field, while the specific value of the dimensionless non-linearity
parameter $f_{\rm{NL}}$ depends on the assumed scenario \citep[see,
e.g.,][]{bartolo2004}. Note that in the literature there are two
conventions for equation (\ref{FNL}), based on the LSS and the CMB,
respectively \citep[see a discussion in][]{matarrese2008}. Here we
follow the first one, for which $\Phi$ is the gravitational potential
linearly extrapolated to $z=0$.

It is worth stressing that detailed second-order calculations of the
evolution of perturbations from the inflationary period to the present
time show that equation (\ref{FNL}) is not generally valid.  In fact
the quadratic, NG contribution to the gravitational potential should
be represented as a convolution with a kernel $f_{\rm NL}({\bf x},{\bf
y})$ rather than a product \citep[see, e.g.,][]{bartolo2005}.  Since
in this paper we will consider models for which $|f_{\rm NL}| \gg 1$,
all space-dependent contributions to $f_{\rm NL}$ can be safely
neglected and $f_{\rm NL}$ can be approximated by a constant.  In
particular the simulations that we will analyse here assume $f_{\rm
NL}= \pm 100, \pm 500, \pm 1000$.  We notice that owing to the
smallness of $\Phi$, the contribution of non-Gaussianity implied by
these values of $f_{\rm NL}$ is always within the percent level of the
total primordial gravitational potential, and does not appreciably
affect the linear matter power spectrum.

We notice that the considered range is larger than that allowed by
present CMB data. In fact, the analysis performed by
\cite{komatsu2008} on the WMAP 5-year maps finds $9 < f_{\rm NL} <
111$ for the 95 per cent confidence level. Our choice for the $f_{\rm
NL}$ range is motivated by two different reasons.  First, the LSS
provides observational constraints which are {\it a priori}
independent of the CMB.  Second, $f_{\rm NL}$ is not guaranteed to be
scale independent, while the LSS and CMB probe different
scales. Indeed some inflationary scenarios do predict large and
scale-dependent $f_{\rm NL}$ \citep[see,
e.g.,][]{chen2005,shandera2006}.

\section{N-body simulations}\label{sect:nbody}

The set of 7 simulations presented here \citep[see
also][]{grossi2007,hikage2008} is characterized by different values of
$\fnl$, and adopts the same so-called `concordance' $\Lambda$CDM model
consisting of a flat universe dominated by a dark energy component
given in the form of a cosmological constant $\Lambda$.  The relevant
cosmological parameters match those determined using the WMAP
first-year data \citep{spergel2003}, i.e. $\Omega_{m0}=0.3$ for the
matter density parameter, $\Omega_{\Lambda0}=0.7$ for the $\Lambda$
contribution to the density parameter, $h=0.7$ for the Hubble
parameter (in units of $100$ km s$^{-1}$ Mpc$^{-1}$).  The initial
power spectrum, which adopts a cold dark matter (CDM) transfer
function, has a spectral index $n=1$ and its amplitude guarantees that
the r.m.s. of density fluctuations measured in spheres of radius $8
h^{-1}$ Mpc is $\sigma_{8}=0.9$.  We recall that the recent analysis
of the WMAP 5-year data \citep{komatsu2008,dunkley2008} provided a
slightly different set of best-fitting cosmological parameters. In
particular, they suggest smaller values for both $\sigma_{8}$ and
$\Omega_{\rm m0}$.  Adopting the first set of parameters rather than
the second does affect the outcome of some statistical analyses
presented in this work.  However, it does not influence the general
results that focus on the {\it relative} differences between models
characterized by a different degree of primordial non-Gaussianity.

All N-body simulations have been performed at the CINECA
Supercomputing Centre (Bologna) and at the `Rechenzentrum der
Max-Planck-Gesellschaft' (Garching) by using the publicly available
code GADGET-2 \citep{springel2005}, which is a more flexible and
efficient version of the numerical code GADGET \citep{springel2001},
thanks to the improved parallelization strategy.  The gravitational
acceleration is computed using the tree method, in combination with a
particle-mesh (PM) scheme for long-range force. In this work we follow
the evolution of collisionless dark matter particles, only.  Therefore
we are not making use of the hydrodynamic capability of the numerical
code.

In all experiments we loaded a computational box of (500 Mpc/$h)^3$
with $800^{3}$ particles: the corresponding particle mass is then
$m=2.033 \times 10^{10} h^{-1} M_\odot$.  The gravitational force has
a Plummer-equivalent softening length of $\epsilon_{l}=12.5 h^{-1}$
kpc.  The grid used by GADGET in the PM scheme was set to have
$1024^3$ nodes.  The runs produced 31 outputs from the initial
redshift and the present time; the redshifts of the last 25 outputs
have been fixed in order to have a comoving distance of $250 h^{-1}$
Mpc (i.e. half box) between them: this will allow to easily build
past-light cone realizations for different complementary studies
already in progress.

\subsection{Setting non-Gaussian initial conditions}\label{sect:IC}

The most critical step in performing the simulations is setting up
suitable NG initial conditions without modifying the power spectrum of
density fluctuations, $P(k)$, which has to be identical to that of the
Gaussian case.

To do that we start by performing a random realization of a Gaussian
gravitational potential $\Phi_{\rm L}$ characterized by a power-law
spectrum $P(k) \propto k^{-3}$. This is done in Fourier space using a
$1024^3$ grid, by applying the standard technique based on the
combination of a random phase with an amplitude extracted from a
Rayleigh distribution corresponding to the desired $P(k)$. This
gravitational potential is then inverse-Fourier transformed back to
real space to include the NG term: $\Phi_{\rm NL}= f_{\rm{NL}}
\left(\Phi_{\rm L}^2 - \langle\Phi_{\rm L}^2\rangle \right)$. As a
final step we go back to $k$-space to account for the CDM matter
transfer function.  This procedure, which improves the original one
proposed by \cite{moscardini1991}, guarantees that different NG models
with different values of $\fnl$ have the same linear power spectra,
all consistent with that of the Gaussian case, as we have checked.
Besides the Gaussian case with $\fnl=0$, we have considered six
different NG models characterized by $\fnl=\pm 100, \pm 500,
\pm 1000$.

The initial conditions for our N-body simulations are then obtained by
using the resulting gravitational potential to perturb a uniform
distribution of particles according to the Zel'dovich
approximation. In order to avoid applying this technique beyond its
range of validity, the initial redshift $z_i$ is fixed in such a way
that the maximum (absolute) value for the density fluctuations on the
grid is smaller than 0.5.  As a result the redshift at which we start
the simulation depends on the $\fnl$ value and varies in the range
$[73,100]$ with the highest redshift corresponding to the largest
$\fnl$ (absolute) value.  In order to avoid spurious effects due to
the non perfect isotropy of a grid, we prefer to perturb a glass-like
distribution, obtained as a result of a N-body simulation where the
sign of gravity has been reversed. As discussed by \cite{white1994},
this technique avoids the presence of preferred directions in the
cubic grid, so the initial distribution retains the uniformity
feature. For this purpose, we replicate 8 times a $100^3$ glass file
along each Cartesian axis, exploiting the periodic conditions.

\begin{figure} 
\psfig{figure=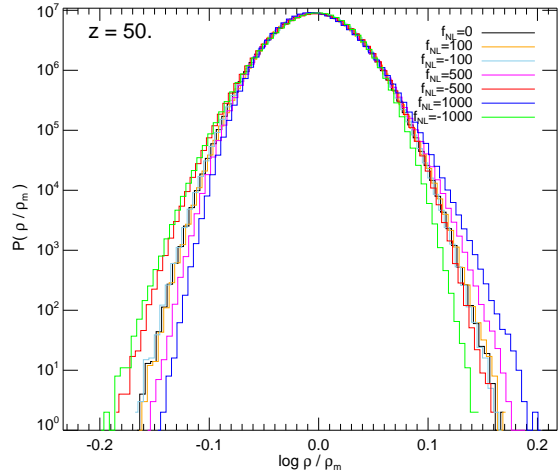,width=0.45\textwidth}
\caption{The PDF for the different models (as
indicated in the plot) as computed at the earliest common redshift
($z=50$). }
\label{fig:PDFinit} 
\end{figure}

In Fig.~\ref{fig:PDFinit} we show the probability density function of
the density field, as measured in the earliest output ($z=50$) common
to all our 7 simulations.  It is computed on a $512^3$ grid, using the
Triangular Shape Cloud (TSC) scheme \citep[see,
e.g.][]{hockney1988}. From the plot it is evident that the larger the
value of $\fnl$, the more pronounced is the high-density tail. This
reflects on the values of the skewness of the density distribution
$\langle \delta^3 \rangle$, which ranges at $z=50$ between $1.3\times
10^{-5}$ for $\fnl=-1000$ and $1.4\times 10^{-4}$ for
$\fnl=1000$. Note that the positive values of the skewness in models
with negative $\fnl$ are produced by the first stages of the
gravitational evolution. In fact, taking into account our sign choice
in the relation between potential and density, we expect a negative
(positive) skewness for the PDF when a negative (positive) $\fnl$ is
used. We have checked this by analyzing the simulated density field at
the redshift corresponding to our initial conditions.  We find that at
$z_i$ the skewness is $-3.9\times 10^{-6}$ and $1.2\times 10^{-5}$,
for $\fnl=-1000$ and $\fnl=1000$, respectively. The small asymmetry
in the two values of the skewness is partially due to sample variance
and also to the fact that the two quantities refer to different epoch
 since $z_i=96.7$ for $\fnl=-1000$ and $z_i=100.8$ for
  $\fnl=+1000$. The redshift evolution of the probability distribution
function of density fluctuations and its moments will be discussed in
detail in Sections~\ref{sect:pdf} and \ref{sect:moments}.

\begin{figure*} 
\psfig{figure=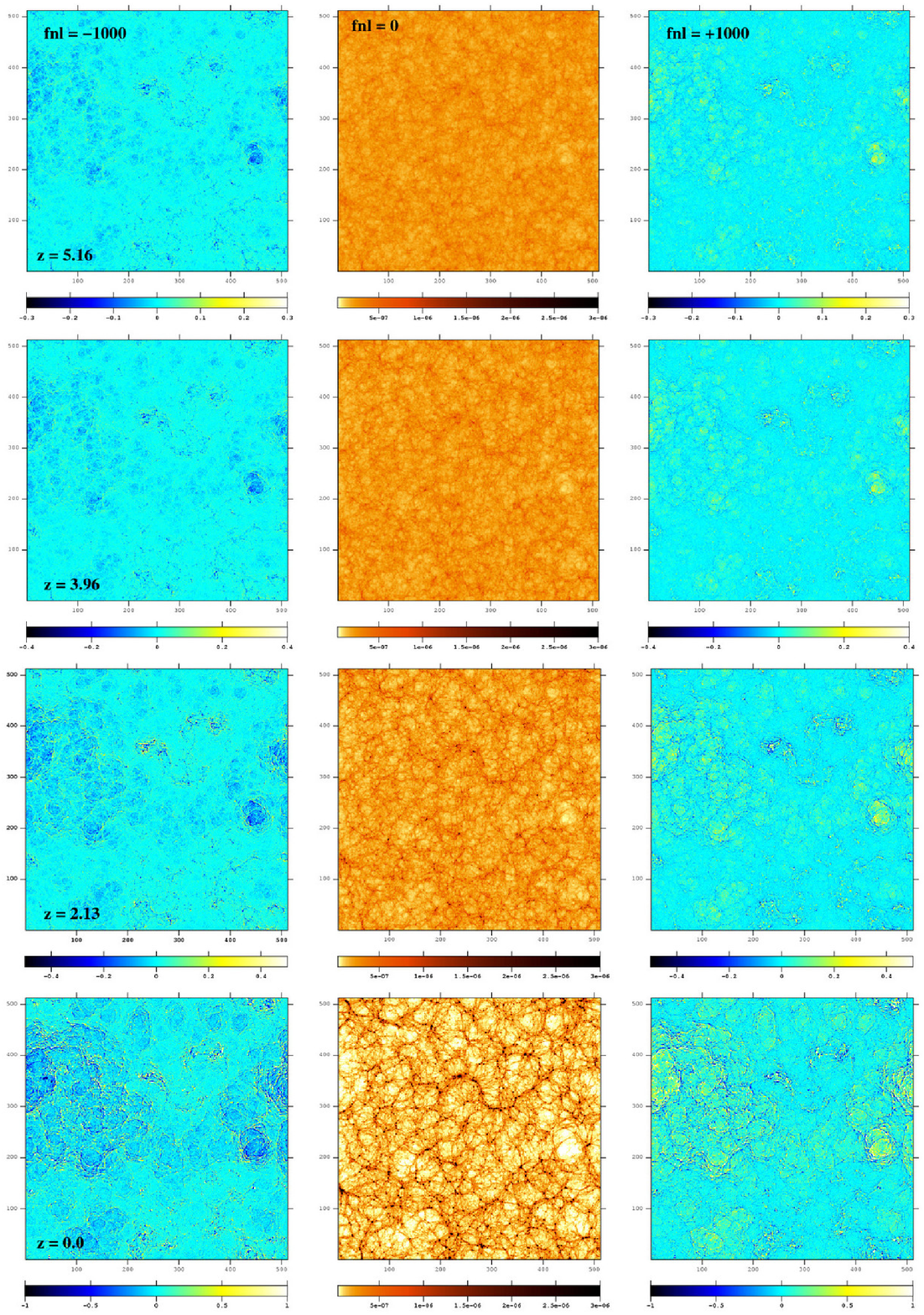,width=0.9\textwidth}
\caption{
Slice maps of the simulated mass density fields at $z=5.15$, $z=3.96$,
$z=2.13$ and $z=0$ (rows from top to bottom).  The side length is
$500h^{-1}$ Mpc, while the thickness is $31.25h^{-1}$ Mpc. The panels
in the middle column show the logarithm of the projected density.  The
left and right panels are the relative residuals for the $f_{\rm
NL}$=$\pm 1000$ runs (equation [\ref{eq:nresidual}]). Each panel has
the corresponding colour bar; note that for the sake of clarity the
considered ranges are different from redshift to redshift.  }
\label{fig:slices} 
\end{figure*} 

\section{Results}\label{sect:evolution}

In this section we investigate the evolution of the dimensionless
power spectrum $\Delta^2(k)$, and the one-point distribution function
of the mass overdensity $p(\delta,R_s)$, focusing on its high-order
moments and low-density tails.

\subsection{Redshift evolution of clustering}\label{sect:clustering}

A first visual impression of how the value of $\fnl$ affects the
clustering evolution can be gained by looking at
Fig. \ref{fig:slices}, where we show maps of the mass density field at
different redshifts ranging from $z=5.16$ and $z=0$.  Each slice has a
side length of $500h^{-1}$ Mpc and a thickness of $31.25h^{-1}$ Mpc.
The central column presents the results for the Gaussian simulation
(i.e. $\fnl=0$), while the relative residuals for $f_{\rm NL}=-1000$
and $f_{\rm NL}=1000$ are shown in the corresponding panels of the
left and right columns.  The residuals are calculated at each pixel as
\begin{equation}
\Delta \rho_x = (\rho_x-\rho_0)/\rho_0\ , 
\label{eq:nresidual}
\end{equation}
where $\rho_x$ represents the density for the NG map with $f_{\rm
NL}=x$ and $\rho_0$ refers to the Gaussian field.  All density fields
in the plots have evolved from the same random realization of the
underlying linear gravitational potential, irrespective of the
statistics.  For this reason, the maps of the residuals
show similar features but with opposite signs.

As a first quantitative statistical test, we consider the
dimensionless power spectrum $\Delta^2$, defined as the contribution
to the variance of the fractional density per unit $\ln k$:
\begin{equation}
\Delta^2(k)\equiv \frac{d\sigma^2}{d \ln k}=\frac{1}{2\pi^2} k^3 P(k)\ .
\end{equation}
In the previous formula $P(k)$ is the matter density power spectrum.

\begin{figure}
\begin{center}
\includegraphics[width = 0.45\textwidth]{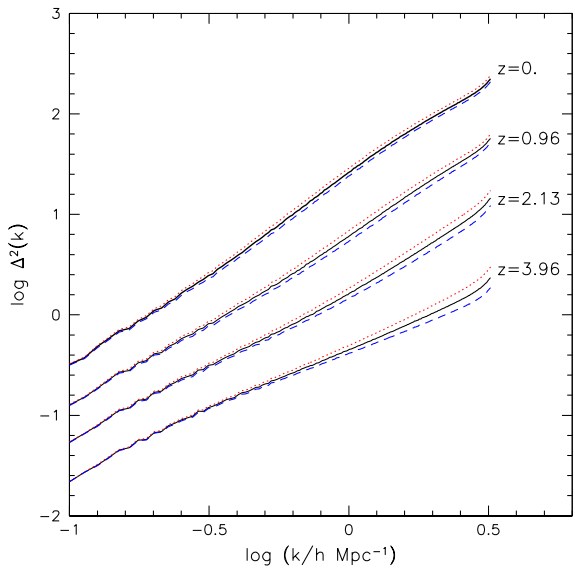}
\caption{The redshift evolution of the dimensionless power spectrum 
$\Delta^2$.  For clarity, only the non-linear scales are
shown. Different lines refer to models with different primordial
non-Gaussianity: $\fnl=0$ (solid line), $\fnl=1000$ (dotted line),
$\fnl=-1000$ (dashed line).  }
\label{fig:delta2}
\end{center}
\end{figure}

The effect of non-Gaussianity can be appreciated in
Fig.~\ref{fig:delta2} where we show $\Delta^2(k)$ computed at four
different redshifts, ranging from $z=3.96$ to $z=0$.  For the sake of
clarity we present the results only for the scales which are in the
non-linear regime (i.e. for $k> 0.1$ $h$/Mpc). Moreover we show only
the two most extreme NG models ($\fnl=\pm 1000)$, which are compared
to the Gaussian reference one.

\begin{figure}
\begin{center}
\includegraphics[width = 0.45\textwidth]{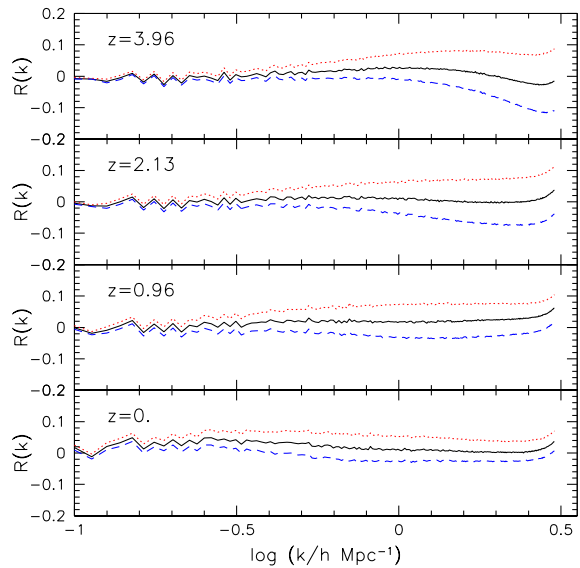}
\caption{The logarithmic deviation $R(k)\equiv \log
\Delta^2_{\rm sim}(k)-\log \Delta^2_{\rm th}(k)$ is shown for the same
models and redshifts presented in Fig.~\ref{fig:delta2}. Here
$\Delta^2_{\rm th}(k)$ is computed using the analytic expression of
{\protect \cite{smith2003}}, which is valid only for the Gaussian
case.  }
\label{fig:errek}
\end{center}
\end{figure}

As expected, at a given redshift, a positive value of $\fnl$ tends to
increase the power at small scales, anticipating the formation of
cosmic structures with respect to the Gaussian case.  On the opposite,
negative $\fnl$ values delay structure formation.
  Fig.~\ref{fig:delta2} shows that the effect is very small at $z=0$
  and increases with the redshift.  However, even at $z=3.96$ the
  largest deviations from the Gaussian case at $k=0.5$ $h$/Mpc do not
  exceed the 30 per cent level, i.e. well below those found in the
  negative tail of the PDF, as we will show in the next Section. In
order to better quantify this effect, in Fig.~\ref{fig:errek} we
present, for the same models and redshifts shown in
Fig.~\ref{fig:delta2}, the logarithmic deviation, $R(k)$, of the
measured $\Delta^2_{\rm sim}(k)$ from the analytic expression
$\Delta^2_{\rm th}(k)$ that \cite{smith2003} have obtained from the
analysis of a large library of N-body simulations.  We point out that
this analytic model can be applied only to models with Gaussian
initial conditions.  Indeed in the simulation with $\fnl=0$ we find
$R(k)\equiv \log \Delta^2_{\rm sim}(k)-\log \Delta^2_{\rm
  th}(k)<0.01$.  In the two NG models with $\fnl=\pm 1000$ the
dimensionless power spectrum deviates from the Gaussian case (and from
the \cite{smith2003} predictions) at $k\sim 1$ $h$/Mpc by
  $|R|\sim 0.04$ at $z=3.96$, $|R|\sim 0.05$ at $z=2.13$ and $z=0.96$,
  and $|R|\sim 0.03$ at $z=0$.  Notice that at redshifts $z\ga 4$
(not shown in the plot), the differences between the dimensionless
power spectra computed for the Gaussian and NG models decrease because
non-linear effects on the scale of interest become negligible at large
redshifts.  This result suggests that the best redshift range to spot
primordial non-Gaussianity is $2\lesssim z \lesssim 4$.

\subsection{Probability distribution 
function of density fluctuations}\label{sect:pdf}

The simplest way to characterize the cosmological density field is its
one-point PDF, i.e. the probability of a density fluctuation $\delta$
measured at the generic position, on a suitable smoothing scale,
$R_s$.  If the PDF of the primordial density field is Gaussian then it
remains Gaussian as long as density fluctuations evolve in the linear
regime.  When $|\delta|\sim 1$, non-linear evolution strongly modifies
the Gaussian shape of the original PDF.  This picture has been
confirmed by the results of different N-body simulations performed in
the cold dark matter framework \citep[see, e.g.][and references
therein]{kayo2001}. These numerical experiments show that the evolved
PDF is well approximated by a lognormal distribution:
\begin{equation}
P_{LN}(\delta)=\frac{1}{\sqrt{2\pi S^2}} \exp \left[
-\frac{\left[\ln(1+\delta)+S^2/2\right]^2} {2S^2}\right]
\frac{1}{1+\delta}\ ,
\label{eq:LN}
\end{equation}
where the parameter $S$ is related to the variance $\sigma^2\equiv
\langle \delta^2 \rangle$ of the density fluctuation field:
$S^2=\ln(1+\sigma^2)$.  The validity of this result has been
thoroughly checked in the weakly non-linear regime but also seems to
apply well in the non-linear regime, at least for $\sigma \la 4$ and
$\delta \la 100$. This result is further corroborated by the
statistical analysis of the three-dimensional distribution of galaxies
\citep[see, e.g.,][]{croton2004,marinoni2005}.  A 
discussion of the possible role of the lognormal distribution for
cosmological structure formation can be found in \cite{coles1991}.

Here we extend the previous analysis by investigating to what extent
the lognormal model can be successfully applied to describe the
evolved PDF in scenarios with primordial non-Gaussianity.  To do this
we have computed the mass density field of our simulations using the
TSC scheme to distribute each particle mass at the points of a regular
grid.  To study the clustering evolution we have considered different
outputs corresponding to different redshifts ($z=3.96, 2.13, 0.96,
0$). Moreover, since the non-linear scale increases with time we have
also considered two different grids: a finer one consisting of $512^3$
points, corresponding to a smoothing radius of $R_s\sim 0.98$ Mpc/h,
suitable for the early stages of the evolution, and a coarser one,
with $128^3$ points ($R_s\sim 3.91$ Mpc/h) for the low-redshift
outputs.  The PDF at the very early times ($z=50$) has been already
shown in Fig.~\ref{fig:PDFinit}.  It is evident that deviations from
Gaussianity are very tiny even for the most extreme NG models, as
quantified by the low values of the skewness.

\begin{figure*}
\begin{center}
\includegraphics[width = 0.45\textwidth]{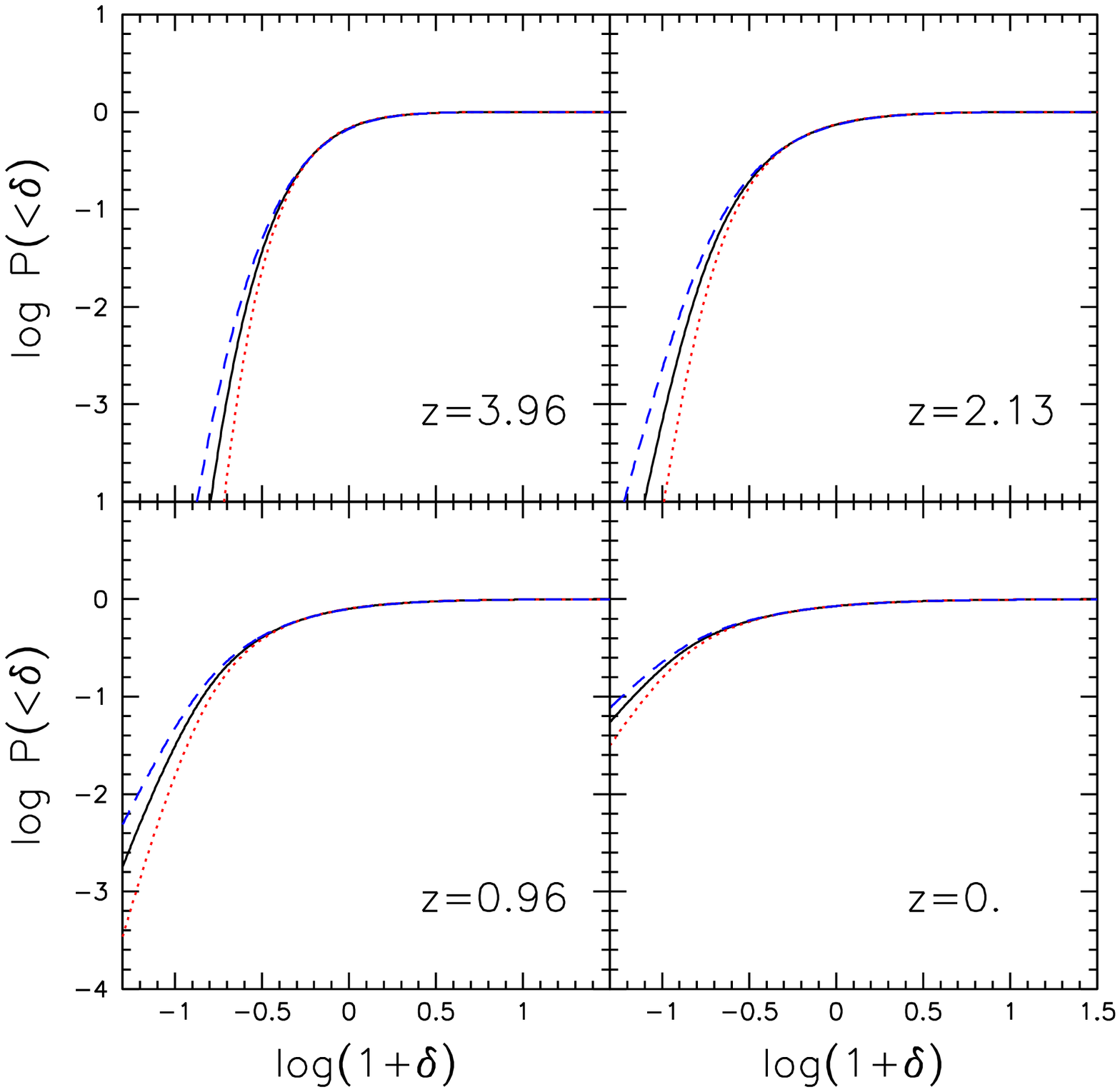}
\includegraphics[width = 0.45\textwidth]{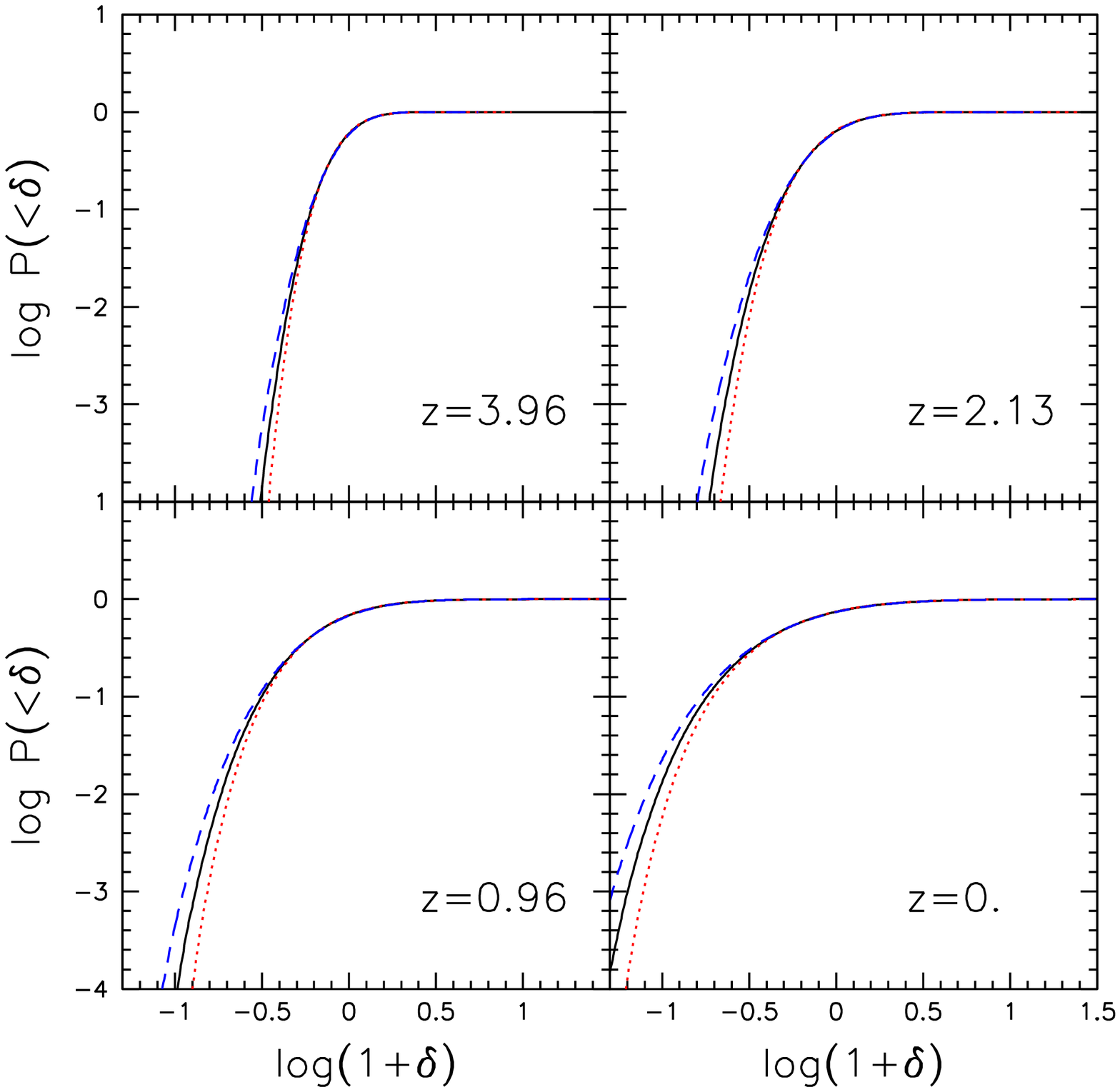}
\caption{ The redshift evolution of the cumulative PDF of the
  density field smoothed on two different scales. Panels on the left:
  smoothing radius $R_s\sim 0.98$ Mpc/h.  Panels on the right:
  smoothing radius $R_s\sim 3.91$ Mpc/h.  The different panels refer
  to the four redshifts explored, indicated in the plots.  Different
  line-styles refer to models with different primordial
  non-Gaussianity: $\fnl=0$ (solid line), $\fnl=1000$ (dotted line),
  $\fnl=-1000$ (dashed line).  }
\label{fig:PDF1}
\end{center}
\end{figure*}

 In Fig.~\ref{fig:PDF1} we show the cumulative PDFs  at four different 
epochs.
We prefer to show the cumulative rather than the differential PDF
since, for a given overdensity threshold, the difference in the PDFs quantifies the difference in the volume occupied by structure below threshold, i.e. 
a quantity that can be potentially measured from real datasets.
It is evident from the plots in Fig.~\ref{fig:PDF1}
that the largest differences
between models are in the low-density tail corresponding to
$\log(1+\delta)\la -1$, as expected from a visual impression of
  Fig. \ref{fig:slices} in which the most prominent features in the
  map of the residuals occur in regions that are below average density
  in the Gaussian case (shown in the central panels).  These residuals
  are seen already at $z=5.16$ and their magnitude increases with
  time.  
 Negative $\fnl$ values increase the probability of
underdense regions with respect to the Gaussian case.  The opposite
holds true for positive $\fnl$. On the contrary, in the high-density
tail, the differences between the models are much less evident. The
differences are more noticeable at $z=2.13$ and decrease both at
higher and lower redshifts, as already pointed out.  

Finally, departures from the  Gaussian case   are more evident
when the density field is smoothed on a small scale 
($R_s \sim 0.98$ Mpc/h in the plots on the 
left hand side of  Fig.~\ref{fig:PDF1}),  but can still be appreciated
when smoothing on a scale of $\sim 3.91$ Mpc/h (shown in the panels on
the right).   Considering the density maps with $R_s \sim 3.91$ Mpc/h,
commonly adopted in similar analyses, and an underdensity threshold 
 $\delta=-0.9$, a  value close to the mean void underdensity
in numerical simulations \citep{colberg2008}, we find that at $z=0$
only 1.3 per cent of the volume is below threshold in the Gaussian case.
The difference with the NG models is striking, since the volume reduces by 
a factor 2.3 in the case of $\fnl=1000$ and increases by  
$\sim 70$ per cent when $\fnl=-1000$. At higher redshifts the differences further increase. 
With the same overdensity threshold we find that if  $\fnl=1000$ 
the volume occupied by regions with $\delta<0.9$ is 
a factor of $\sim 6.5$ smaller than in the Gaussian case while
if  $\fnl=-1000$ the volume is $\sim 17$ times  larger.
One should note, however, that the volume fraction below 
this density threshold rapidly decreases with the redshift.
At $z=0.9$ this fraction is only 0.007 per cent for $\fnl=0$.
The fact that at all redshifts departures from
Gaussianity preferentially occur in low-density environments strongly
suggest that dedicated statistical tests related to the void
probability could be successfully applied to spot primordial
non-Gaussianity. 
It is worth pointing out that this is a promising observational test
also at the present epoch since the volume occupied by voids, in which 
the NG features are still well preserved, increases with time. 
 
In order to check if a lognormal distribution provides a good
description of the evolved PDFs even in the NG case, we compute the
logarithmic deviation, $\Delta \log P\equiv
\log PDF - \log P_{LN}$,  where the reference lognormal  $ P_{LN}$
is computed from equation (\ref{eq:LN}), adopting for the variance
$\sigma^2$ the value computed directly from the N-body outputs.  We
notice that the value of the mass variance $\sigma^2$ that
we measure in the Gaussian simulation agrees, at the few percent
level, with the theoretical predictions of \cite{peacock1996} and
\cite{smith2003}, hence confirming the results of \cite{kayo2001}.

\begin{figure*}
\begin{center}
\includegraphics[width = 0.45\textwidth]{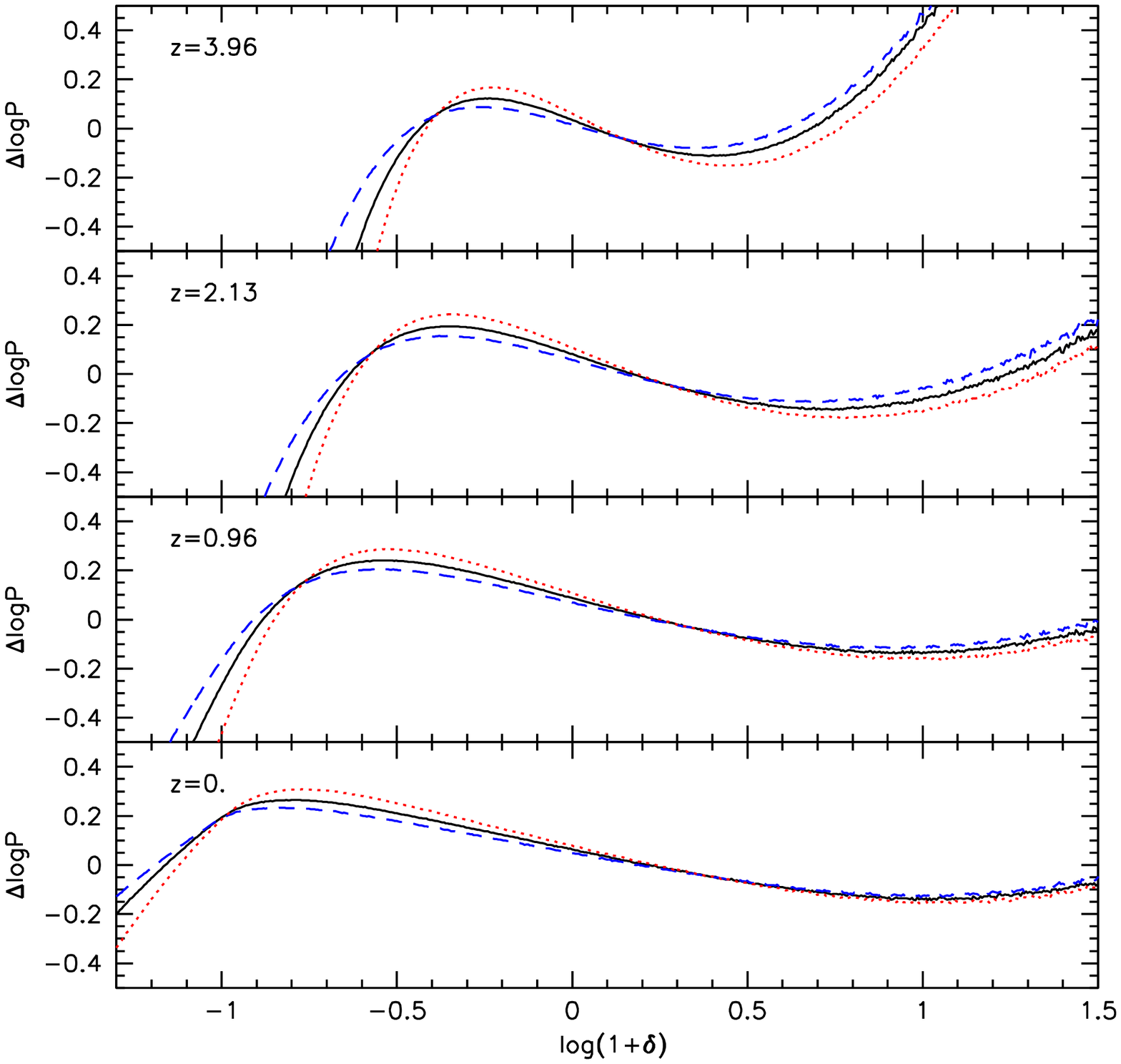}
\includegraphics[width = 0.45\textwidth]{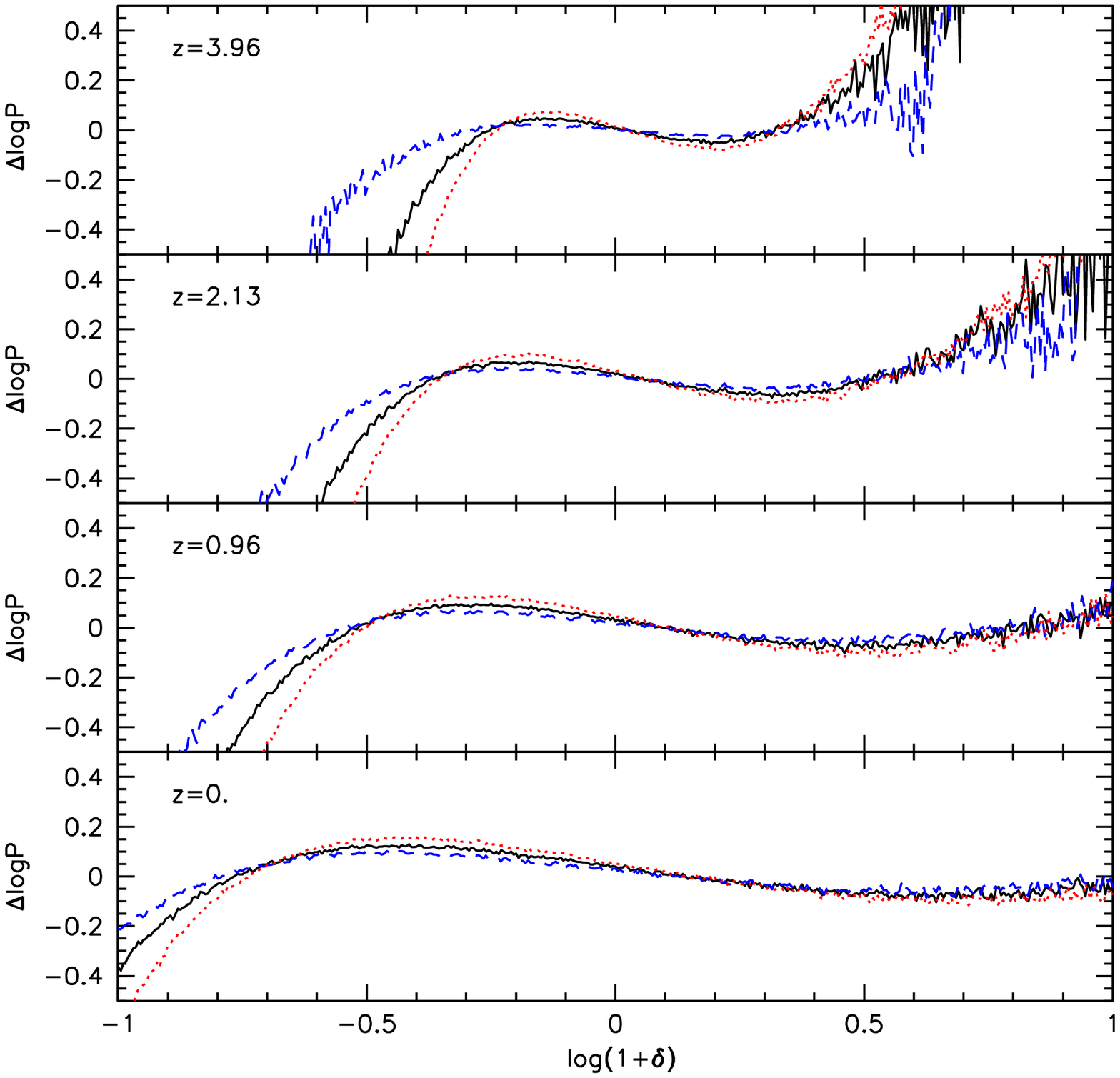}
\caption{The logarithmic deviation of the  
PDF from a lognormal distribution, $\Delta \log P$, is shown for the
same models and redshifts presented in Fig.~\ref{fig:PDF1}.  Results
for smoothing radii $R_s\sim 0.98$ and $R_s\sim 3.91$ Mpc/h are
displayed in the left and right panels, respectively.  Different lines
refer to models with different primordial non-Gaussianity: $\fnl=0$
(solid line), $\fnl=1000$ (dotted line), $\fnl=-1000$ (dashed line).
}
\label{fig:PDF2}
\end{center}
\end{figure*}

The results are displayed in Fig.~\ref{fig:PDF2} for the same models,
redshifts and smoothing radii presented in Fig.~\ref{fig:PDF1}.  In
general a lognormal distribution provides a reasonable fit to the
evolved PDFs. The match is very good at intermediate overdensities
($-0.2\la \log(1+\delta) \la 0.6$) and improves when the smoothing
length increases.  The deviation $\Delta \log P$ increases towards
both the high and the low-density tails. This trend is systematic and
independent of the primordial non-Gaussianity.  In particular the
lognormal distribution tends to over(under)-estimate the probability
of the under(over)-dense regions, and therefore cannot describe the
statistical properties of both voids and highly-clustered regions.
 We recall that analytic theoretical predictions for the PDF,
  different from the lognormal model, which agree at few per cent
  level in the void regime have been very recently proposed by
  \cite{lam2008}.

\subsection{High-order moments}\label{sect:moments}

A way to quantify the PDF properties is to measure its high-order
moments. Rigorously speaking all moments have to be specified to fully
describe a distribution function. However, for practical purposes a
reasonable characterization of the bulk of the PDF can be achieved
using the small-order moments up to the fourth.  Therefore, here we
consider the second, the third and the fourth-order moments of the
distribution, namely its variance $\sigma^2\equiv \langle
\delta^2\rangle$, skewness $\gamma\equiv \langle
\delta^3\rangle$ and kurtosis $\kappa\equiv \langle \delta^4\rangle$.

\begin{figure}
\begin{center}
\includegraphics[width = 0.45\textwidth]{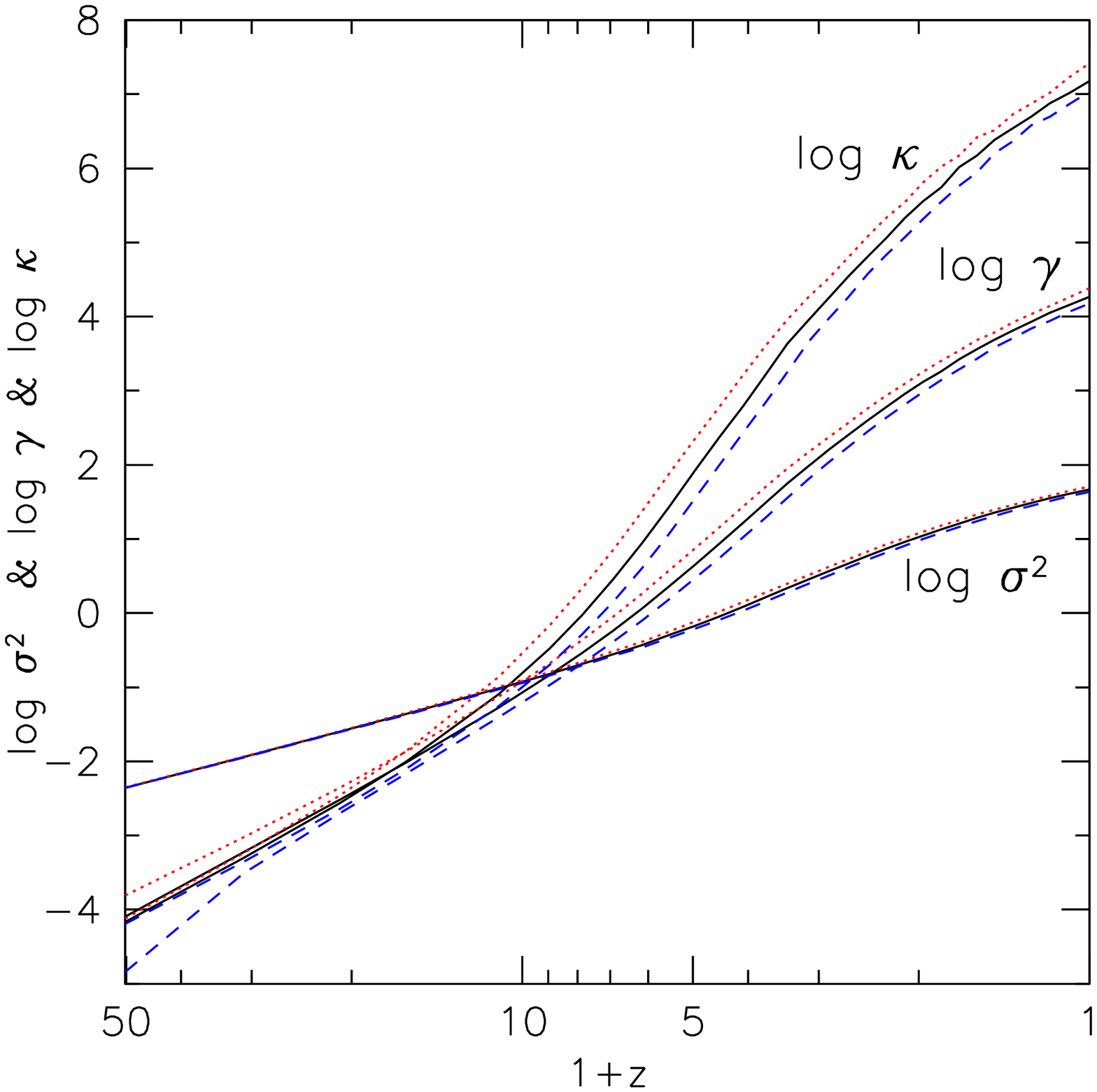}
\caption{The redshift evolution of variance $\sigma^2$, skewness $\gamma$ 
and kurtosis $\kappa$. Labels on the plot identifies the different
moments.  Different line-styles refer to models with different
primordial non-Gaussianity: $\fnl=0$ (solid line), $\fnl=1000$ (dotted
line), $\fnl=-1000$ (dashed line).  }
\label{fig:moments}
\end{center}
\end{figure}

The gravity-driven clustering evolution amplifies these moments. Their
growth rate, however, does not depend on the level of primordial
non-Gaussianity.  This is evident from Fig.~\ref{fig:moments}, where
we show the redshift evolution of variance, skewness and kurtosis
relative to the PDF measured on the finer ($512^3$) grid, i.e. to a
smoothing radius $R_s\sim 0.98$ Mpc/h.  For the sake of clarity and to
avoid overcrowding, we display the results for the Gaussian case
($\fnl=0$) and for the two most extreme NG models ($\fnl=\pm 1000$)
only.  The evolution of each moments is very similar in the three
models, characterized by different line-styles. This is particularly
evident for the variance, $\sigma^2$, and reflect the fact that the
power spectra of the various models remain almost identical throughout
the evolution.  Focusing on $\gamma$ we note that NG models with
positive (negative) $\fnl$, characterized by a positive (negative)
primordial skewness, have values of $\gamma$ that are systematically
larger (smaller) than the Gaussian case  by about 25 per cent . This difference, originated
from the different initial conditions, is kept at the same level
through the evolution history. Indeed, to a first approximation, the
gravitational evolution is expected to couple high-order moments to
the variance (see below), the evolution of which has only a weak
dependence on $\fnl$, as witnessed by the fact that the evolution of
$P(k)$ depends only weakly on $\fnl$ (see Section
\ref{sect:clustering}).  This is also confirmed by the analysis of the
other simulations of NG models with $|\fnl|<1000$ (not shown).  These
same considerations apply to the kurtosis, $\kappa$ for which, however,
deviations from the Gaussian case at $z=0$ amount to $\sim 50$ per cent.

Starting from Gaussian density fluctuations, perturbation theory shows
that gravitational instability leads to hierarchical scaling relations
between the variance of the PDF and its higher-order moments.  For
this reason it is useful to repeat the previous analysis using two
different quantities: the normalized skewness, defined as, $S\equiv
\gamma/\sigma^4$ and the normalized kurtosis, $K\equiv (\kappa -3
\sigma^4)/\sigma^6$.  Predictions of the perturbation theory by
different authors have been successfully tested using N-body
simulations \citep[see, e.g.,][and references
therein]{lucchin1994}. The results, however, revealed that the values
of $S$ and $K$ depend on the shape of the primordial power spectrum
and on the shape and scale $R$ of the smoothing filter.  For instance,
with a Gaussian filter and a cold dark matter power spectrum, typical
values are $S\sim 3$ and $K\sim 16$ \citep{kayo2001}.  These values
also agree with the predictions of the lognormal model described by
equation (\ref{eq:LN}), for which
\citep{kayo2001}
\begin{equation}
S(R)=3+\sigma^2(R)
\label{eq:s_r}
\end{equation}
and
\begin{equation}
K(R)=16+15 \sigma^2(R)+6\sigma^4(R) +\sigma^6(R)\ .
\label{eq:k_r}
\end{equation}

We recall that the hierarchical scaling relations are a prediction
derived under the assumption of Gaussian fluctuations. For this
reason, possible deviations from the theoretical predictions have to
be regarded as a signature of primordial non-Gaussianity.  This is in
essence the so-called {\rm skewness test} suggested by
\cite{coles1991} and \cite{silk1991}, according to which departures
from a constant value $S$ on scales in which fluctuations are still
evolving in the weakly non-linear regime provide a powerful test for
NG models, as demonstrated by the analysis of first-generation NG
simulations \citep{coles1993}.

\begin{figure}
\begin{center}
\includegraphics[width = 0.45\textwidth]{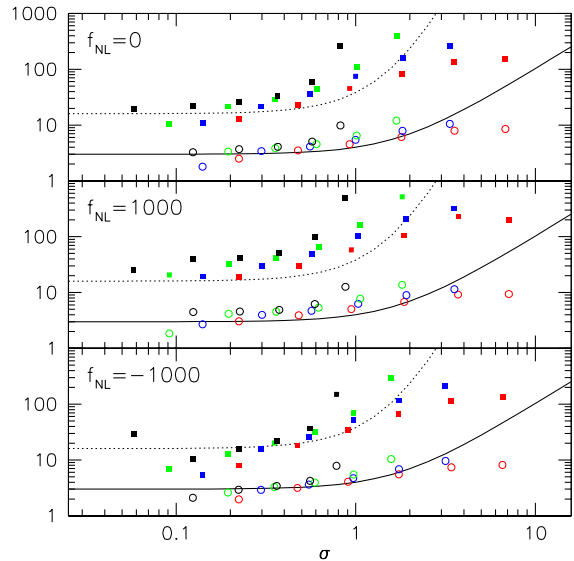}
\caption{
The normalized skewness $S$ (open circles) and the normalized kurtosis
$K$ (filled squares) of the simulated density field as a function of
$\sigma$. The corresponding theoretical values predicted by equations
(\ref{eq:s_r}) and (\ref{eq:k_r}) are shown with a solid and a dashed
line, respectively.  Different panels refer to models with different
primordial non-Gaussianity.  Top: Gaussian model $\fnl=0$.  Middle: NG
model with $\fnl=+1000$ .  Bottom: NG model with $\fnl=-1000$ .
Moments have been computed using different smoothing radii
[$R_s=(1,2,4,8,16,32)\times \Delta x$, with $ \Delta x\sim 0.98$
Mpc/h].  The colours of the symbols correspond to measurements at
different redshifts: $z=3.96$ (red), $z=2.13$ (blue), $z=0.96$ (green)
and $z=0$ (black).  }
\label{fig:mom2}
\end{center}
\end{figure}

We have repeated the same test using both the normalized skewness $S$
and kurtosis $K$.  The results are shown in Fig.~\ref{fig:mom2} in
which both quantities are shown as a function of the
r.m.s. $\sigma$. The different panels refer to different numerical
simulations and compare the measured $K$ (filled squares) and $S$
(open circles) with theoretical predictions of the lognormal
  model for which $S(R)$ and $K(R)$ are provided by equations
  (\ref{eq:s_r}) and (\ref{eq:k_r}) (solid and dashed curves,
  respectively).  Each panel combines results obtained at different
redshifts ($z=3.96$, $z=2.13$, $z=0.96$ and $z=0$) and using different
smoothing radii ($R_s=(1,2,4,8,16,32)\times \Delta x$, with $ \Delta
x\sim 0.98$ Mpc/h).  The magnitude of both moments increases with
$\sigma$, independently of the value of $\fnl$.  Focusing on the top
panel, we notice that in the low-$\sigma$ regime (i.e. $\sigma \la 1$)
$S(\sigma)$ flattens but is not quite constant, which would be at
variance with the predictions of the hierarchical model.  However, we
attribute this mismatch to a spurious effect deriving from the use of
different smoothing scales.  A similar trend is also seen in the two
NG cases in the same low-$\sigma$ range. We conclude that, given the
smallness of the effect, it is difficult to disentangle the signature
of primordial non-Gaussianity from the spurious signal produced by the
varying smoothing radii, hence reducing the power of the skewness
  test. Using the normalized kurtosis has two potential advantages.
  First, the signal is more than one order of magnitude larger than
  the normalized skewness. And second, the differences between
  Gaussian and NG cases in the same $\sigma$-range are significantly
  larger.  However, as shown by \cite{kayo2001}, kurtosis has larger
  error than skewness.  Therefore to decide which statistics is more
  sensitive to primordial non-Gaussianity one should perform an
  accurate error analysis, which is beyond the scope of this work.
  Such analysis would also allow to estimate and correct for
  systematic effects induced by the finite size of the computational
  box although in our case, given the large size of the computational box, 
  the correction factor, estimated according to \cite{kayo2001}, is very
  small and can be safely neglected.

The plot confirms that in the weakly non-linear regime a lognormal
model predicts fairly well the normalized high-order
moments. Discrepancies become very significant during the non-linear
evolution, i.e. when $\sigma \ge 3$ in the $S$ case and when $\sigma
\ge 1$ in the $K$ case. We note that in the high-$\sigma$ regime the
lognormal model systematically over-predicts the amplitude of the
normalized moments in both the Gaussian and the NG models.

\section{Discussion and conclusions}\label{sect:conclusions}

In this work we have investigated the properties of the mass density
field extracted from the N-body experiments of NG scenarios presented
in \cite{grossi2007}. These models are physically motivated in the
context of the inflationary theory, and cover a large range of
primordial non-Gaussianity: $-1000 < f_{NL} < 1000$. Moreover the
state-of-the art (mass, force and spatial) resolution of these
simulations allows us to study the evolution of the primordial
non-Gaussianity on both linear and non-linear scales, from very high
redshifts to the present epoch.

Our results show that clustering properties are little affected by
primordial non-Gaussianity, which is soon erased by the effect of
non-linear evolution. Indeed, the power spectrum of our most extreme
models matches the Gaussian case within 20-30 per cent.  On the
contrary, the one-point PDF of the mass density field represents a
promising tool to distinguish among competing NG models. In fact,
while in the high-density tail the primordial signal is quickly
obliterated by non-linearly induced non-Gaussianity, in the
low-density regions the primordial non-Gaussianity stands out,
expecially at intermediate ($1\la z \la 2$) redshifts. The differences
are preserved up to the present epoch.

This result suggests that statistics related to the void distribution
could provide a powerful tool to detect primordial non-Gaussianity.
Unfortunately, studying low-density environments is challenging from
both a practical and a theoretical point of view.  First of all, a
theoretical difficulty is represented by the fact that no such
statistical tool has been proposed in the context of NG models.
Furthermore, even in Gaussian scenarios, there is no unique way of
estimating a well-defined statistic like, for example, the void
probability function.  In this respect, significant progress has been
recently made to define efficient and reliable void finding algorithms
\citep[see for example the comparison presented in][and references
therein]{colberg2008}.  Finally, all statistics aimed at probing
low-density regions are notoriously difficult to apply because of the
sparseness of the data.  However, this problem might be overcome by
exploiting quasar high-resolution spectra thanks to the intrinsic
relation between the properties of the void in the three-dimensional
mass distribution and the "flux-voids" \citep[defined as the connected
regions in the one-dimensional flux distribution above the mean flux
level; see][]{viel2008}.  We plan to investigate this issue in detail
by means of dedicated, high-resolution hydrodynamical simulation of NG
models.

Using spectra of distant objects to spot voids allows, in principle,
to study their properties and therefore to investigate non-Gaussianity
at all redshifts.  At high $z$ this method will complement other
statistics which are also effective in detecting NG features, like the
occurrence of rare events represented by the abundance of massive
clusters \citep[see, e.g.,][]{grossi2007}, the topology of the mass
distribution, that has been investigated by \cite{hikage2008}.
Finally, in this paper we have studied the skewness and the kurtosis
of the PDF and found that neither is very sensitive to primordial
non-Gaussianity, even at very high redshifts. In this respect the
bispectrum represents a much more powerful statistical tool that has
been extensively used to constrain the amount of primordial
non-Gaussianity in the CMB \citep[see,
e.g.,][]{komatsu2003,cabella2006,spergel2007,creminelli2007,komatsu2008}.
Also, it has been successfully applied to the existing galaxy surveys
at low redshift to assess the amount of non-linearity in galaxy
biasing \citep{verde2002,gaztanaga2005,pan2005}.  In a future paper we
plan to use our simulations to assess the power of the bispectrum in
high-redshift surveys as a probe of primordial non-Gaussianity
\citep[see][]{sefusatti2007}.

\section*{acknowledgments}
Computations have been performed on the IBM-SP5 at CINECA (Consorzio
Interuniversitario del Nord-Est per il Calcolo Automatico), Bologna,
with CPU time assigned under an INAF-CINECA grant and on the IBM-SP4
machine at the ``Rechenzentrum der Max-Planck-Gesellschaft'' at the
Max-Planck Institut fuer Plasmaphysik with CPU time assigned to the
MPA.  We acknowledge financial contribution from contracts ASI-INAF
I/023/05/0, ASI-INAF I/088/06/0 and ASI I/016/07/0. We thank the
  anonymous referee for her/his comments which allow us to improve the
  presentations of our results. We are grateful to Licia Verde for
useful discussions, and Claudio Gheller for his assistance. We also
thank Gerhard B\"orner for careful reading of the manuscript.

\bibliographystyle{mn2e}
\bibliography{margot_long.bib}

\label{lastpage}
\end{document}